\def\bab{\beta(k)}
\def\cab{\xi(k)}

\def\emkx{e^{-i\vec k\cdot\vec x}}
\def\ess{e^{-i(\vec k\cdot\vec x-k\eta)}}

\def\hab{h_{\alpha\beta}}
\def\eab{\varepsilon_{\alpha\beta}}
\def\sax{(1~+~\mu^{2})\cos 2\phi}

\def\vax{4\mu\sin 2\phi}

\def\fbr{\overline {F}(\eta ^{'},k)}
\def\far{\overline {F}(\eta,k)}
\def\har{H(\eta,k)}
\def\ewp{\eta^{'}}
\def\ewd{\eta^{''}}

\centerline{{\bf POLARIZATION OF THE COSMIC MICROWAVE BACKGROUND}}
\centerline{{\bf IN A REIONIZED UNIVERSE}}
\vskip 1cm
\centerline{\it Marina Gibilisco}\vskip 1mm
\centerline{\it Universit\'a degli studi di Pavia,}\vskip 1mm
\centerline{\it Via Bassi 6, 20127 Pavia, Italy,}\vskip 1mm
\centerline{\it and Universit\'a degli studi di Milano,}\vskip 1mm
\centerline{\it Via Celoria 16, 20133, Milano, Italy.}
\vskip 5mm
{\bf Abstract:}
\vskip 5mm
In this work I study the polarization of the Cosmic Microwave Background (CMB)
induced by a cosmological background of gravitational waves (here GWs), probably
originated during the inflationary epoch.
I discuss the influence of a possible reionization of the Universe, which
should happen at late times ($z\sim 20\div 30$): I show
that, in the presence of such a reionization, we have 
a remarkable enhancement of the present polarization of the CMB.
I also point out the r\"ole of a background of gravitational waves
of large wavelength, re-entering the horizon very late,
in producing the anisotropy and the polarization of the CMB through the
Sachs-Wolf effect.
Then, I use the standard formalism of the Stokes parameters
to describe the properties of the CMB photons and I study their evolution
with the time, the energy and the spatial coordinates through the radiative
Boltzmann transfer equation. I solve such an equation in the form of 
a second-kind Volterra integral equation, by using the analytical
method of the iterated kernels: this particular method can be successfully 
applied when we have an exponential kernel, as in the case when the Universe 
knew a fast and late reionization, due, for instance, to the Hawking 
evaporation of primordial black holes. 
\vskip 7mm
{\bf 1. THE RADIATIVE TRANSFER EQUATION AND THE SACHS-WOLFE EFFECT.}
\vskip 7mm
The equation which expresses the evolution of the Stokes parameters
with the time, the energy and the space coordinates is the 
Boltzmann radiative transfer equation $^{1}$:
$$
\Big(~{\partial \vec n\over \partial \eta}~+~\gamma^{\alpha}~
{\partial\vec n \over
{\partial x^{\alpha}} }~\Big)~+~{\partial\nu \over {\partial\eta}}~
{\partial\vec n\over{\partial\nu}}~=~{\sigma_{T}~N_{e}~R(\eta)~\over 4\pi}~
\times
\Big [~-4\pi\vec n~+~\int^{1}_{-1}~\int^{2\pi}_{0}~\vec n~P(\mu,\phi,\mu ',
\phi ')~d\mu ' ~d\phi '~\Big].\eqno(1.1)
$$
\noindent ($\eta$ is the comoving time,
defined as $\int [dt/R(t)]$, $R(t)=$ being the scale factor of the Universe).

Here, the variables are the components of a 4-vector 
$ n_{\alpha}~=~ n_{\alpha}(\theta,\phi,\nu)$, which is
a function of the polar angles $(\theta, \phi)$ and of the photon frequency;
these components are the four Stokes parameters which describe the polarization
status of the radiation $^{1}$:
$$
n_{\alpha}~\equiv~(I_{l},~I_{r},~U,~V).\eqno(1.2)
$$
$I_{l},~I_{r}$ are the left and right intensities of the radiation, while $U$
and $V$ respectively represent the linear and the circular polarization 
of the CMB photons.
In the Chandrasekhar formalism $^{1}$ one has:
$$
I~=~I_{l}+I_{r}~~~~~~~~~~~~~~~~~~~~~~~~~Q~=~I_{l}-I_{r}.\eqno(1.3)
$$
In eq. (1.1) $\gamma^{\alpha}$ are the components of an unit vector in 
the propagation direction of the photons (in particular, the $\hat z$ 
axis is taken parallel to the propagation direction of the gravitational 
waves we consider), $\sigma_{T}=6.65~\times~10^{-25}$
$cm^{2}$ is the Thomson scattering cross-section, $R(\eta)$ is 
the scale factor, $\mu~=~\cos \theta$, $P(\mu,\phi,\mu ',\phi ')$ 
is the Chandrasekhar scattering matrix and $N_{e}$ is the comoving number 
density of the free electrons: such a variable introduces in the Stokes 
parameters evolution an important dependence on the ionization history 
of the Universe. 
In fact, the Thomson scattering of an anisotropic radiation (as the CMB is)
gives it a linear polarization $^{2}$: clearly, in a reionized Universe,
the Thomson scattering processes are effective for a longer time, thus
one obtains an enhancement of the resulting CMB polarization.

The relevant term in the left-hand side of eq. (1.1) is the third one; 
in fact, it expresses a very peculiar effect due to 
a cosmological background of gravitational waves,
known as Sachs-Wolfe effect $^{3}$. The presence of a ripple in the 
space-time metric produces a shift in the photons frequencies: in fact,
a time-dependent perturbation in the metric, i.e. a gravitational wave, 
determines the non-conservation of the energy along the 
photons sight-line, thus causing a frequency shift:
$$
{\partial\nu\over\partial\eta}~=~{\nu\over 2}~{\partial \hab\over\partial\eta}
\gamma^{\alpha}\gamma^{\beta}.\eqno(1.4)
$$
If one writes the perturbation of the metric tensor as a plane wave
$$
\hab~=~h~\ess\eab,\eqno(1.5)
$$
with
$$
\eab\gamma^{\alpha}\gamma^{\beta}~=~(1~-~{\cos\theta}^{2})\cos 2\phi,\eqno(1.6)
$$
the frequency shift becomes
$$
{\partial\nu\over\partial\eta}~=~{\nu\over 2}~\emkx~\Big({\partial\over
\partial\eta}h~e^{ik\eta}\Big)~(1~-~\mu^{2})~\cos 2\phi.\eqno(1.7)
$$
Thus, the Sachs-Wolfe term obviously influences the evolution of the 
Stokes parameters and, as a consequence, 
there is a polarization induced in the CMB by cosmological
gravitational waves.
\vskip 7mm
{\bf 2. THE SECOND-KIND VOLTERRA EQUATION}
\vskip 7mm
Here I will briefly recall the standard formalism of ref. $^{4}$,
useful to turn the transfer equation (1.1) into 
a second-kind Volterra equation.

By using two new variables, $\alpha$ and $\beta$, and two new vectors:
$$
\vec a~=~{1\over 2}~(1~-~\mu^{2})\cos 2\phi 
\left (
\matrix {1, &1, &0\cr}
\right ),\eqno(2.1)
$$
$$
\vec b~=~{1\over 2}
\left (
\matrix {\sax, &-~\sax, &\vax\cr}
\right ),\eqno(2.2)
$$
the vector $\vec n$ is written in the form
$$
\vec n~=~n_{0}~\left (
\matrix {1, & 1, &0 \cr}
\right )~+~\ess~(\alpha\vec a~+~\beta\vec b).\eqno(2.3)
$$
In eq. (2.3) $\vec n_{0}~=~(n_{0},n_{0},0)$ is the unperturbed solution one has
in absence of gravitational waves.

By expressing $I,~ Q,~ U$ as functions of these variables, one obtains
two coupled, first order differential equations $^{4}$:
$$
\dot\beta~+~[q(\eta)~-~ik\mu]~\beta=~F(\eta,k),\eqno(2.4)
$$
$$
\dot\xi~+~[q(\eta)~-~ik\mu]~\xi=~H(\eta,k),\eqno(2.5)
$$
where $q(\eta)~=~\sigma_{T}N_{e}R(\eta)$ depends, through the electron 
density $N_{e}$, on the ionization history of the Universe.

The right-hand sides of eqs. (2.4) and (2.5) are
$$
F(\eta,k)~=~{3q(\eta)\over 16}~\int ^{1}_{-1}d\mu ^{'}~
\Bigg[~(1~+~\mu ^{'2})^{2}~\beta (\mu^{'},k)~-~{1\over 2}\xi (\mu^{'},k)~
(1~-~\mu ^{'2})^{2}~\Bigg],\eqno(2.6)
$$
$$
H(\eta,k)~=~D(k)k~\Bigg[~{j_{2}(k\eta)\over k\eta}~\Bigg ].\eqno(2.7)
$$
Eq. (2.6) comes from the calculation of the scattering integral 
in eq. (1.1), while $H(\eta,k)$ is given by the time derivative of the 
metric tensor perturbation, calculated by taking into account the 
particular gravitational waves spectrum predicted by the theory $^{5,6}$.
The form of their spectrum is expressed by the factor $D(k)$, which 
is strongly model-dependent: the simplest choice, when quantum gravity effects
are not involved, is $^{7}$
$$
D(k)~=~\sqrt{16\pi G H^{2}(k)};\eqno(2.8)
$$
in eq. (2.8) $H$ is the curvature of the de Sitter Universe at the epoch when 
the GWs enter the horizon.

More complex forms for $D(k)$, referring to the various kinds of inflationary
Universe, can also be considered: here, for simplicity, I will discuss
the simplest case only.

By substituting the variables $\beta$ and $\xi$ in eq.(2.6), 
formally obtained by eqs. (2.4) and (2.5) $^{4}$, the 
resulting equation is a 2-nd kind Volterra integral equation which reads:
$$
\far~=~{3q(\eta)\over 16}~\Bigg\{~\int^{\eta}_{0}~d\eta^{'}~\fbr~
K_{+}(\eta,\ewp,k)
-{1\over 2}~\int^{\ewp}_{0}~d\ewd~H(\ewd,k)~e^{-\tau(\ewd)}~K_{-}(\ewp,\ewd,k)
\Bigg\},\eqno(2.9)
$$
where
$$
K_{\pm}(\eta,\ewp,k)~=~\int^{1}_{-1}~d\mu~(1~\pm~\mu^{2})^{2}~
e^{ik\mu(\eta-\ewp)}\eqno(2.10)
$$
and 
$$
\tau(\eta)~=~\int^{1}_{\eta}~q(\eta^{'})~d\eta^{'}.\eqno(2.11)
$$
is the optical depth.
\vskip 7mm
{\bf 3. THE SOLUTION WITH AN EXPONENTIAL KERNEL:}

{\bf THE METHOD OF THE ITERATED KERNELS}
\vskip 7mm
In Ref. $^{8}$ I discussed a mixed analytical-numerical method to solve 
the radiative transfer equation, based on the resolvent method $^{9}$;
such a method can be used when the kernel of the equation has a 
polynomial form. The interested 
people can find a detailed discussion of this technique in the quoted
reference. Here I want rather to discuss a totally analytical way to
solve the transfer equation, based on the properties of the Volterra
integral equations. The method of the iterated kernel is a very simple and 
elegant way to solve a Volterra
equation, but it can be used in few cases only: the fundamental task is to
find a resolvent function by writing a
convergent series built up with the kernel functions, after they have been
convoluted and integrated. For a detailed discussion of this method 
see refs. $^{9,10}$; here, I will briefly recall the main results
of my analysis, see ref. $^{10}$.

Look now at the general form of a Volterra equation:
$$
f(x)~=~G(x)~+~\int^{x}_{0}K(x,x')~f(x')~dx'.\eqno(3.1)
$$
Here $K(x,x')$ is the kernel and $G(x)$ is the source function.
Starting from eq. (3.1), the iterated kernel is given by $^{9}$:
$$
K_{m}(x,x')~=~\int^{x}_{x'}K(x,t)K_{m-1}(t,x')dt.\eqno(3.2)
$$
If the successive approximations of the kernel obtained 
in such a way really converge,
the integral equation can be written in the form 
$$
f(x)~=~G(x)~+~\sum^{\infty}_{m=1}~\lambda^{m}~\int~K_{m}(x,x')G(x')dx'.
\eqno(3.3)
$$
By defining the resolvent as
$$
\Gamma(x,x';\lambda)~=~\sum^{\infty}_{m=1}~\lambda^{m-1}K_{m}(x,x'),
\eqno(3.4)
$$
the final solution is $^{9}$:
$$
f(x)~=~G(x)~+~\lambda~\int~G(x')~\Gamma(x,x';\lambda)~dx'.\eqno(3.5)
$$
The applicability of this method clearly requires the convergence of the series
in eqs. (3.3) and (3.4); moreover, the integral (3.2) should involve quite 
simple, analytical functions. 

An exponential kernel satisfies these conditions very well:
in our specific case, the kernel can be put in an 
exponential form if the function $q(\eta)$ (see eq. (2.11)),
specifying the ionization history of the Universe, has such a form.
As I proved in $^{11,12}$ a reionization of the Universe 
induced by the Hawking evaporation of Primordial Black Holes (PBHs)
has just an exponential behaviour.

With this choice, the kernel of the Volterra equation can be approximated by
$$
S(k,\eta,\ewp)~=~q(\eta)K_{+}(k,\eta,\ewp)~\sim~
exp(c_{1})~exp(c_{2}k(\eta-\ewp)).\eqno(3.6)
$$
The parameters $c_{1}$ and $c_{2}$ are obtained by a low $\chi^{2}$
fit, whose results are shown in fig. 1 for the case
$z=20,~x=1$, for which $c_1~=~5.22$, $c_2~=~-18.42$.

Finally, the original Volterra equation (2.9) can be turned $^{10}$ into a
very simple, linear, first order differential equation with a 
boundary condition $\overline {F}(k,0)=G(k,0)$; the 
final solution is
$$
\overline{F} (k,\eta)~=~G'(k,\eta)~+~{3\over 32}~{D(k)\over k}
\lambda ~{e^{a}\over b}\Big[~
exp((\lambda+c_{2}k)\eta)(exp(b\eta)-1)~\Big],\eqno(3.7)
$$
where $a$ and $b$ are two constants coming from the calculation of the 
primitives.
\vskip 7mm
{\bf 4. THE CALCULATION OF THE CMB POLARIZATION}
\vskip 7mm
The exact solutions of eqs. (2.4) and (2.5) are obtained by
substituting in eq. (2.9) the functions $\far$ and 
$\har$ determined through the analytical solution of the 
Volterra equation previously found. 
Then, the present CMB polarization and anisotropy, induced by 
a superposition of $k$-fixed GWs, are given by $^{10}$:
$$
P^{2}(k)~=~\int^{1}_{-1}~|\bab|^{2}~\big[(1~+~\mu^{2})^{2}~+~4\mu^{2}\big]~d\mu,
\eqno(4.1)
$$
$$
A^{2}(k)~=~\int^{1}_{-1}~|\cab|^{2}~(1-\mu^{2})^{2}~d\mu.\eqno(4.2)
$$
The normalization is carried out by defining the polarization degree as follows:
$$
p(k)~=~{\sqrt{\overline {Q}^{2}~+~\overline {U}^{2}}\over \sqrt{\overline 
{I}^{2}}};\eqno(4.3)
$$
$$
\overline {I}^{2}(k)~=~|\alpha(k)|^{2}~(1-\mu^{2})^{2}~+~4I_{0}^{2}~+~
4I_{0}~|\alpha(k)|(1-\mu^{2})^{2};\eqno(4.4)
$$
here
$$
\overline {I}_{0}~=~{1\over {exp(h\bar\nu / k_{B}T)~-~1}}
\eqno(4.5)
$$
is the unperturbed intensity; I take also
$$
h\overline\nu~\sim~10^{-13}~GeV,~~~~~~~~~~~~~~~~~k_{B}T~\sim~2.35\times
10^{-13}~GeV.\eqno(4.6)
$$
In the same way, the normalized anisotropy is defined as:
$$
a^{2}(k)~=~{\int^{1}_{-1}~|\cab|^{2}~(1-\mu^{2})^{2}~d\mu \over 
\overline{I}^{2}(k)},\eqno(4.7)
$$
where 
$\overline{I}^{2}(k)$ is still given by eq. (4.4).
\vskip 5mm
In fig. 2 I showed the polarization peak obtained from the solution
of the transfer equation in the case of a standard ionization
history of the Universe $^{8}$; in figs. 3 and 4 I showed my results
respectively in the case of a reionization which happens at $z=10$
and at $z=20$. I plotted the natural logarithm of the polarization degree
vs. the wave number of the considered gravitational waves.
In fig.2 the maximum of the polarization corresponds to a wave-number $k\sim 15$,
i.e. to gravitational waves re-entering the horizon at the end of the 
recombination epoch: their effect is dominant.
The peak is the direct consequence of some mathematical suppressions:
for $\eta~<~0.03$, the source function $G(\eta,k)$ is strongly suppressed by 
the exponential function;
the physical reason for this suppression is 
that the polarizing effect cannot be relevant because the Thomson 
scattering has a low efficiency and thus the perturbations are overally damped. 
As a consequence, GWs re-entering the horizon too early are unimportant 
in order to produce a CMB polarization.

For $0.03<\eta<0.076$, Thomson scattering just begins to 
have some importance, but the polarizing effect of GWs having 
a wavenumber $k~=~15$ is dominant, thus masking that of GWs of different
$k$. Therefore, the polarization is relevant only near $\eta~=~0.066$.
Finally, if none reionization is considered, for $\eta~\geq~0.076$ the 
ionization degree is very small: as a result, the polarizing effect is 
negligible because we have no more free electrons which scatter the CMB photons.

As one can see from figs. 3 and 4, a sharp secondary peak appears also 
at smaller $k$ ($k=4$), in the case of a total reionization coming at $z=20$.
That represents the enhancement of the CMB polarization in a 
reionized Universe: in this case, a polarization for the CMB photons 
should be observable at angular scales near $9^{\circ}$, slightly smaller
than the ones tested by COBE. On the contrary, for a standard ionization
history, one should expect to see the CMB polarization only at angular
scales near $2^{\circ}$, thus requesting very high precision measurements.
\vskip 7mm
{\bf ACKNOWLEDGMENTS}
\vskip 7mm
I am especially grateful to the University of Pavia, which financially
supports my work and gives me also the technical way to perform my
research. Thank you also to all the people of the Astrophysical Section
of the University of Milano and, in particular, to my Tutor there,
Silvio Bonometto, for all the things he taught me and for his 
friendly and continuous support. Thank you also to my friends, Emma
Ficara and Francesco Cavaliere, to whom I am in debt for many things.

\vskip 5mm
{\bf REFERENCES}
\vskip 5mm

\item {1} Chandrasekhar, S., ``Radiation Transfer'', Dover, New York, (1960).

\item {2} Rees, M. J., {\it APJ}, {\bf 153}, L1, (1968).

\item {3} Sachs R.K., Wolfe, A.M., {\it APJ}, {\bf 147}, 73, (1967).

\item {4} Basko, M.M., Polnarev, A.G.: {\it Sov. Astr.}, {\bf 24}, 268, (1980).

\item {5} Polnarev, A.G., {\it Sov. Astron.}, {\bf 29}, 607, (1985).

\item {6} Bardeen, J.M., Bond, J.R., Salopek, D.S., {\it Phys. Rev.}, {\bf D40},
1753, (1989).

\item {7} Starobinskii, A.A., {\it Sov. Astr. Lett.}, {\bf 11}, 133, (1985).

\item {8} Gibilisco, M.:  {\it Astroph. and Space Sc.}, {\bf 235}, 75,
(1996).

\item {9} Mikhlin, S.G., Integral Equations and their applications,
Oxford Pergamon Press, (1964);

\item {} Elsgolts, P: ``Equazioni differenziali e calcolo 
delle variazioni'', Editori Riuniti, (1990).

\item {10} Gibilisco, M.:  {\it Int. Journ. of Mod. Phys.},  {\bf 10A}, 3605, 
(1995).

\item {11} Gibilisco, M.:  {\it Int. Journ. of Mod. Phys.},  {\bf 11A}, 5541, 
(1996).

\item {12} Gibilisco, M.: {\it Int. Journ. of Mod. Phys.}, {\bf 12A}, 4167,
(1997).

\end